\documentclass[conference]{IEEEtran}
\IEEEoverridecommandlockouts
\usepackage{cite}
\usepackage{amsmath,amssymb,amsfonts}
\usepackage{algorithmic}
\usepackage{algorithm}
\usepackage{bm}
\usepackage{braket}
\usepackage{graphicx}
\usepackage{textcomp}
\usepackage{xcolor}
\usepackage{multirow}
\usepackage{siunitx}
\def\BibTeX{{\rm B\kern-.05em{\sc i\kern-.025em b}\kern-.08em
    T\kern-.1667em\lower.7ex\hbox{E}\kern-.125emX}}

\usepackage{booktabs}
\usepackage{siunitx}
\begin{document}

\title{ 
Scaling Sample-Based Quantum Diagonalization on GPU-Accelerated Systems using OpenMP Offload
}
\author{
\IEEEauthorblockN{Robert Walkup}
\IEEEauthorblockA{\textit{IBM Research} \\
\textit{IBM T.J. Watson Research Center}\\
Yorktown Heights, NY, USA \\
walkup@us.ibm.com}
\and
\IEEEauthorblockN{Juha J\"aykk\"a}
\IEEEauthorblockA{\textit{Advanced Micro Devices, Inc.} \\
\textit{AMD Silo AI}\\
Cambridge, UK\\
Juha.Jaykka@amd.com}
\and
\IEEEauthorblockN{Igor Pasichnyk}
\IEEEauthorblockA{\textit{Advanced Micro Devices, Inc.} \\
\textit{Global Center of Excellence}\\
Munich, Germany \\
Igor.Pasichnyk@amd.com}
\and
\IEEEauthorblockN{Zachary Streeter}
\IEEEauthorblockA{\textit{Advanced Micro Devices, Inc.} \\
\textit{Artificial Intelligence Group}\\
Austin, TX, USA \\
Zachary.Streeter@amd.com}
\and
\IEEEauthorblockN{Kasia \'{S}wirydowicz}
\IEEEauthorblockA{\textit{Advanced Micro Devices, Inc.} \\
\textit{HPC \& AI Software Enablement} \\
Austin, TX, USA \\
Kasia.Swirydowicz@amd.com}
\and
\IEEEauthorblockN{Mikko Tukiainen}
\IEEEauthorblockA{\textit{Advanced Micro Devices, Inc.} \\
\textit{AMD Silo AI}\\
Turku, Finland\\
Mikko.Tukiainen@amd.com}
\and
\IEEEauthorblockN{Yasuko Eckert}
\IEEEauthorblockA{\textit{Advanced Micro Devices, Inc.} \\
\textit{Research and Advanced Development}\\
Bellevue, WA, USA \\
Yasuko.Eckert@amd.com}
\and
\IEEEauthorblockN{Luke Bertels}
\IEEEauthorblockA{\textit{Quantum Information Science Section} \\
\textit{Oak Ridge National Laboratory}\\
Oak Ridge, TN, 37831, USA \\
bertelslw@ornl.gov}
\and
\IEEEauthorblockN{Daniel Claudino}
\IEEEauthorblockA{\textit{Quantum Information Science Section} \\
\textit{Oak Ridge National Laboratory}\\
Oak Ridge, TN, 37831, USA \\
claudinodc@ornl.gov}
\and
\IEEEauthorblockN{Peter Groszkowski}
\IEEEauthorblockA{\textit{National Center for Computational Sciences} \\
Oak Ridge, TN, 37831, USA \\
groszkowskip@ornl.gov}
\and
\IEEEauthorblockN{Travis S. Humble}
\IEEEauthorblockA{\textit{Quantum Science Center} \\
Oak Ridge, TN, 37831, USA \\
humblets@ornl.gov}
\and
\IEEEauthorblockN{Constantinos Evangelinos}
\IEEEauthorblockA{\textit{IBM Research} \\
\textit{IBM T.J. Watson Research Center}\\
Yorktown Heights, NY, USA \\
cevange@us.ibm.com}
\and
\IEEEauthorblockN{Javier Robledo-Moreno}
\IEEEauthorblockA{\textit{IBM Research} \\
\textit{IBM T.J. Watson Research Center}\\
Yorktown Heights, NY, USA \\
j.robledomoreno@ibm.com}
\and
\IEEEauthorblockN{William Kirby}
\IEEEauthorblockA{\textit{IBM Research} \\
\textit{IBM T.J. Watson Research Center}\\
Yorktown Heights, NY, USA \\
William.Kirby@ibm.com}
\and
\IEEEauthorblockN{Antonio Mezzacapo}
\IEEEauthorblockA{\textit{IBM Research} \\
\textit{IBM T.J. Watson Research Center}\\
Yorktown Heights, NY, USA \\
mezzacapo@ibm.com}
\and
\IEEEauthorblockN{Antonio Córcoles}
\IEEEauthorblockA{\textit{IBM Research} \\
\textit{IBM T.J. Watson Research Center}\\
Yorktown Heights, NY, USA \\
adcorcol@us.ibm.com}
\and
\IEEEauthorblockN{Seetharami Seelam}
\IEEEauthorblockA{\textit{IBM Research} \\
\textit{IBM T.J. Watson Research Center}\\
Yorktown Heights, NY, USA \\
sseelam@us.ibm.com}
}

\maketitle
\thispagestyle{plain}
\pagestyle{plain}
\begin{abstract}
Hybrid quantum–HPC algorithms advance research by delegating complex tasks to quantum processors and using HPC systems to orchestrate workflows and complementary computations. Sample-based quantum diagonalization (SQD) is a hybrid quantum–HPC method in which information from a molecular Hamiltonian is encoded into a quantum circuit for evaluation on a quantum computer. A set of measurements on the quantum computer yields electronic configurations that are filtered on the classical computer, which also performs diagonalization on the selected subspace and identifies configurations to be carried over to the next step in an iterative process.  Diagonalization is the most demanding task for the classical computer.  Previous studies used the Fugaku supercomputer and a highly scalable diagonalization code designed for CPUs. In this work, we describe our efforts to enable efficient scalable and portable diagonalization on heterogeneous systems using GPUs as the main compute engines based on the previous work.

GPUs provide massive on-device thread-level parallelism that is well aligned with the algorithms used for diagonalization.  We focus on the computation of ground-state energies and wavefunctions using the Davidson algorithm with a selected set of electron configurations.  We describe the offload strategy, code transformations, and data-movement, with examples of measurements on the Frontier supercomputer and five other GPU-accelerated systems.  Our measurements show that GPUs provide an outstanding performance boost of order 100x on a per-node basis. This dramatically expedites the diagonalization step—essential for extracting ground and excited state energies—bringing the classical processing time down from hours to minutes. By aligning the classical runtime with quantum execution duration, this approach facilitates the efficient study of complex chemical systems on heterogeneous architectures.
\end{abstract}

\section{Introduction}
Hybrid algorithms that combine quantum computing with high-performance computing (HPC) are opening up new areas for research, where quantum computers handle specific complex tasks, and HPC systems manage workflows and tackle complementary compute-intensive functions. Recent hybrid studies include investigations of electronic structure in quantum chemistry and material science, where the quantum computer generates a collection of sampled states and passes information to the HPC system for further analysis~\cite{robledo2024chemistry, yu2025quantum, shirakawa2025closedloopcalculationselectronicstructure, piccinelli2025quantum, kaliakin2025accurate, shajan2025toward, shajan2025molecular}.  In this work we focus on sample-based quantum diagonalization (SQD), a recently developed hybrid quantum-HPC method designed to use near term quantum computers for electronic structure calculations. This method is an extension of conventional selected configuration interaction (SCI) in quantum chemistry into the quantum-computing domain ~\cite{kanno2023QSCI}. SQD was used with the Fugaku supercomputer to scale up electronic structure calculations to sizes beyond the reach of exact methods ~\cite{robledo2024chemistry, shirakawa2025closedloopcalculationselectronicstructure}. In SQD, information extracted from the molecular Hamiltonian is first encoded into a quantum operator and implemented as a parameterized circuit on a quantum computer. By sampling the resulting quantum state, the algorithm generates a collection of electronic configurations whose probabilities reflect their contributions to the ground or excited states of interest. These sampled electronic configurations are then passed to the HPC system, which filters and ranks them, and then constructs a reduced subspace of the Hilbert space. Potential and current applications of SQD and its variations include the study of the ground- and excited-state properties of metalorganic complexes~\cite{robledo2024chemistry, shirakawa2025closedloopcalculationselectronicstructure, Barison_2025_ext-SQD}, molecules relevant to the description of combustion chemistry present in rocket plumes~\cite{Liepuoniute2025_SQD}, the understanding of hydrogen abstraction processes relevant to the photo-degradation of
composite materials~\cite{smith2025quantumcentricsimulationhydrogenabstraction_SQD}, the study of the structure of small proteins~\cite{shajan2026proteins_SQD}, as well as the description of the ground state properties of impurity models relevant to realistic materials calculations through embedding techniques~\cite{yu2025quantum, sriluckshmy2025ghostgutzwiller_SQD}.

Within this quantum-sampled and classically-filtered subspace, a conventional diagonalization procedure is performed to obtain approximate eigenvalues and eigenvectors of the Hamiltonian. The results of this classical diagonalization not only improve the current estimate of the energy but also determine which configurations should be carried over to the next step in an iterative process~\cite{shirakawa2025closedloopcalculationselectronicstructure}.
In this way, SQD forms an iterative feedback loop between quantum sampling and classical post-processing, progressively refining the relevant subspace until convergence is achieved. This iterative structure allows SQD to balance the strengths of quantum hardware --- in particular, the ability to sample from high-dimensional quantum states --- with the reliability and precision of linear-algebra routines on classical systems.

From a quantum chemistry perspective, the methods described in this work fall under the category of selected configuration interaction, SCI~\cite{holmes2016heat_SCI, zhang2025TrimCI_SCI, holmes2016efficient_SCI, tubman2016deterministic_SCI, sharma2017semistochastic_SCI, Huron_1973_CIPSI}.  In this approach, a finite set of one-electron basis functions referred to as spin-orbitals, is chosen, and the multi-electron wavefunction is represented by a linear combination of anti-symmetrized products of the spin-orbitals, denoted as configurations $\Phi$.  Every term in the expansion corresponds to a particular ``configuration" where each electron is placed in a specific spin-orbital.  Occupation of a given spin-orbital can be represented by a bit-string, where the number of bits is equal to the number of spin-orbitals, and occupation is indicated by a 1 or 0 entry in the bit-string.  A full configuration interaction (FCI) calculation includes all possible configurations, and provides the most accurate  description of the electronic wavefunction within the limits of the finite basis set.  However, the total number of terms in the FCI expansion increases exponentially with the number of spin-orbitals, making that approach impractical in many cases.  The goal in SCI is to include a reduced set of configurations that provide good accuracy while remaining computationally tractable.  In our SQD approach, the quantum computer provides samples of electronic configurations, which are then filtered and refined classically before the diagonalization step.  Addressing large molecules or large basis sets using SQD necessitates large numbers of samples from the quantum computer and correspondingly large numbers of configurations for the classical calculation that follows.  Hence, a highly performant code for the classical component of SQD is of paramount importance, which motivates the present work. 

The dominant computational cost arises from the iterative diagonalization of large Hamiltonian matrices constructed from selected configurations. Methods such as the Davidson algorithm~\cite{davidson1975iterative} are used to solve ground-state energies and wavefunctions within this configuration-based subspace. However, for large molecular systems, each Davidson iteration requires the evaluation of millions to billions of Hamiltonian matrix elements $H_{ij} = \langle \Phi_i | \hat{H} | \Phi_j \rangle$, making the diagonalization step the primary computational bottleneck.

Previous implementations on the Fugaku supercomputer~\cite{fugaku_supercomputer, shirakawa2025closedloopcalculationselectronicstructure,sqd-diag} demonstrated excellent CPU scalability. However, the advent of exascale heterogeneous systems with GPU accelerators presents both opportunities and challenges. GPUs provide massive thread-level parallelism, tens of thousands of concurrent threads, that aligns naturally with the underlying matrix and vector operations. However complex data structures and deeply nested control flow in quantum chemistry codes pose significant porting challenges, requiring careful design of computational kernels, data layouts, and offload strategies.

This work presents a systematic approach to GPU acceleration of the SQD method using OpenMP 5.0+ target offload directives. We focus on portability and maintainability, enabling a single code base to execute efficiently on both CPU-only and GPU-accelerated systems. To support reproducibility and further research, we have open-sourced the full implementation at the link in the footnote~\footnote{https://github.com/AMD-HPC/amd-sbd}. While prior studies discuss the possibility of GPU-accelerated implementations of certain subroutines in SCI methods~\cite{Tubman2020}, to the best of our knowledge, this work shows the first benchmarking study of the GPU implementation of the subspace diagonalization subroutine of SCI or SQD. Our contributions include:

\begin{itemize}
    \item A data-layout strategy that converts nested C++ data structures into flattened, GPU-friendly arrays,
    \item Implementation of a persistent configuration cache that eliminates redundant computations and improves data reuse,
    \item A fully device-resident implementation of all routines to evaluate Hamiltonian matrix elements, and
    \item Comprehensive numerical validation demonstrating agreement with CPU results to within $10^{-10}$ relative error.
\end{itemize}

Performance measurements on the Frontier supercomputer~\cite{frontier_supercomputer_ornl} demonstrate roughly 100$\times$ speedup compared to the CPU implementation~\cite{sqd-diag}, with agreement to within a relative error of $10^{-10}$. This reduces the time-to-solution from hours to minutes for representative molecular systems. Additional benchmarking on newer GPU-accelerated platforms --- including H100, GB200, MI355X, and MI300X --- shows further speedups of 1.8$\times$ to 3$\times$ over Frontier. These performance gains make SQD calculations tractable on moderately sized GPU resources and significantly expand the range of quantum chemistry applications that can be tackled efficiently.

\section{Analyzing CPU code}
Before starting the process of enabling GPUs, considerable effort went into analyzing the CPU code~\cite{sqd-diag}, in order to understand the key data structures and to identify performance-critical code sections.  The CPU code supports two main methods (1) a matrix-free approach where elements of the Hamiltonian matrix must be re-computed during each Davidson iteration, and (2) a method where the Hamiltonian matrix is computed one time and stored, and the stored values are re-used during each Davidson iteration.  The matrix-free method saves memory at the expense of extra computational work, and is generally the method of choice for large matrix dimensions.  Consequently we focused on the matrix-free method.  Several test cases were constructed using the PySCF quantum chemistry package~\cite{Sun2020PySCF} to perform a full configuration interaction calculation using a common basis set, and then select a subset of the electron configurations that had coefficients in the ground-state wavefunction with a magnitude larger than some cutoff value.  This approach yielded sets of test cases~\cite{sqd-diag} where the number of configurations ranged from $6\times10^4$ to $2\times10^9$.  In the matrix formulation, each row of the matrix corresponds to a specific configuration.  

The CPU profiles~\cite{perfetto_tracing_tool, amd_uprof_performance_analysis} showed that at a high level, the key routine is matrix-vector multiplication, where the vast majority of the time is spent re-computing the Hamiltonian matrix elements, $H_{ij}$ as shown in Table~\ref{tab:exclusive_time}.  
In the table, the times in the percent column include only the contribution of code directly in the listed function, not counting function calls to other routines.  For the routines that matter, \texttt{mult()} is the parent routine, which calls \texttt{Hij()} and \texttt{DetFromAlphaBeta()}.  The \texttt{Hij()} routine makes calls to \texttt{OneExcite()} and \texttt{TwoExcite()}, and those make calls to the \texttt{parity()} routine.  There is some time spent directly in the \texttt{mult()} routine.  Other functions contributed very small fractions to the total compute time.

There is abundant parallelism to exploit: each row of the matrix can be independently multiplied with the input vector to yield one element of the product vector.  The code to compute each $H_{ij}$ element is deeply nested and makes extensive use of C++ \textit{std::vector} objects.  

Electron configurations are encoded as bit-strings, where occupation of a given spin-orbital is indicated by a 1 or 0.  The CPU code spends a lot of time analyzing bit-strings to determine which orbitals contribute to a given element of the Hamiltonian matrix.  Eventually a look-up table is used to retrieve integrals that were provided in an input file, and the integral values are added to the matrix element, modulo a sign which must also be computed.  After the value of the matrix element has been computed, the matrix-vector product can be incremented, but that is a very minor step compared to evaluation of $H_{ij}$.  

\begin{table}[h!]
\centering
\caption{Exclusive function time breakdown of the CPU code.}
\begin{tabular}{r l}
\hline
\textbf{Percent Time} & \textbf{Function (exclusive)} \\
\hline
42.88\% & \texttt{sbd::Hij<double>} \\
35.15\% & \texttt{sbd::DetFromAlphaBeta} \\
15.55\% & \texttt{sbd::parity} \\
3.61\%  & \texttt{sbd::mult<double>} \\
2.74\%  & \texttt{sbd::OneExcite<double>} \\
0.01\%  & \texttt{sbd::difference} \\
0.01\%  & \texttt{sbd::GenerateExcitation} \\
0.01\%  & \texttt{sbd::Davidson<double,double>} \\
0.01\%  & \texttt{sbd::ZeroExcite<double>} \\
0.01\%  & \texttt{sbd::InnerProduct<double>} \\
\hline
\end{tabular}
\label{tab:exclusive_time}
\end{table}

The CPU code uses a tensor-product of separate bit-strings for spin-up (alpha) or spin-down (beta) spin-orbitals, and it repeatedly constructs full configuration bit-strings by interleaving the bits of the separate alpha and beta bit-strings.  Additionally, each bit-string is represented as a C++ \textit{std::vector} of 64-bit integer objects, and only a portion of each vector element is used for encoding bits in the bit-string, as specified by a \textit{bit\_length} parameter.

We used hardware counters on CPUs to get a picture of instruction mix and memory-bandwidth utilization (see Appendix A for details).  The overall summary for optimized CPU code was that each CPU core could complete $~2.5$ instructions per clock-cycle, where the instruction mix was $~55\%$ integer, $~26\%$ load-store, $~18\%$ branch, and only $~1\%$ floating-point, and where memory bandwidth utilization was negligible, at $~0.3\%$ of peak.  These characteristics are a consequence of the code required to re-compute each $H_{ij}$ matrix element.  One might be tempted to think that such a code would not perform well on a GPU, because there is no opportunity to use tensor cores or to benefit from high-bandwidth memory.  However, as we will show, GPUs are very effective in driving high instruction throughput for this application.                             

\section{Strategies to develop portable GPU code}

Starting with the existing C++ CPU code, there are several options for enabling GPU acceleration.  First, we noticed from CPU profiles that the vast majority of compute time was in one routine that takes care of matrix-vector multiplication, so a viable approach would be to offload that routine, including the deeply nested functions required to compute matrix elements, to the GPU, and leave the remaining code untouched.  It would be necessary to transfer data to the GPU, including tables of one- and two-electron integrals, and lists that characterize the sparse structure of the Hamiltonian matrix.  Those data items can be transferred one time and remain resident on the GPU for all iterations.  It would also be necessary to transfer the input vector and manage the output vector for each iteration.  The data transfer times per iteration were expected to be small compared to the time required to compute the matrix elements.

An alternative approach would be to move all steps of the Davidson algorithm to the GPU: vector updates, dot-products, norms, and Gram-Schmidt orthogonalization, not just the matrix-vector multiply routine.  Doing so would require changes to the core algorithm, significantly increasing the number of large vectors stored on the GPU, and would also require messaging libraries to support communication using GPU-resident send and receive buffers.  The potential benefit is that moving these steps to the GPU should provide a little more performance as described in a related paper~\cite{horii}.  In this work, we have taken the first approach: offload only the matrix-vector multiply routine.

Having decided on an offload strategy, there are several options for a portable implementation.  These include OpenMP target directives~\cite{openmpoffloadoverview}, re-structuring the code to use the Thrust library~\cite{nvidia_thrust_manual}, or the lower-level HIP interface~\cite{hip}, which allows finer control of GPU code generation.  The CPU code makes extensive use of OpenMP~\cite{dagum1998openmp}, and so using OpenMP target directives was the natural choice.  This approach requires the fewest code changes, and is portable to the widest range of accelerator devices.  OpenMP target directives are used to manage data movement and to offload computational work.

\section{Porting SQD code using OpenMP Target offload directives}

The code provides four ways to distribute work and memory: (1) partition the alpha bit-strings, (2) partition the beta bit-strings, (3) use task-based parallelism, and (4) optionally assign the work of multiplying a given row of the matrix to a rank in the ``row communicator"~\cite{sqd-diag, shirakawa2025closedloopcalculationselectronicstructure}.  In our offload approach, we attempt to preserve all of these options.  The parallel decomposition can be described by four integers (a,b,t,r) reflecting the number of MPI ranks used to partition the alpha and beta bit-strings (a,b), the number of ranks (t) used to concurrently work on computational tasks, and the distribution of rows to ranks (r) in the row communicator.  The number of computational tasks is determined by the way the bit-strings are partitioned, according to a simple formula: $a\times b + a + b$.  Matrix-vector multiplication is done inside a task-loop, and there are three distinct types of tasks dictated by the Slater-Condon rules, where bra and ket configurations differ by: (1) one or two alpha excitations, (2) one or two beta excitations, or (3) one alpha and one beta excitation~\cite{knowles1984new}.  Consequently there are just three doubly-nested loops over alpha and beta bit-strings that require OpenMP target directives to offload the computational work.  We used the ``collapse(2)" clause to enable threading over the combined $alpha \times beta$ index space, and so the number of available threads is equal to the total number of configurations in each of the three doubly-nested loops  - typically numbers of order $10^5$ to $10^7$ threads.  This is more than sufficient to ensure high utilization of all compute units on the GPU.

Rather than attempting to port all relevant code sections in a single pass, we adopted an incremental strategy implementing distinct optimization phases. Each phase builds upon previous work and can be independently validated, facilitating debugging and performance analysis.

\subsection{Persistent Configuration Cache}
\label{sec:cache}
In the CPU performance profiles, a significant fraction of the time was consumed by re-computing configuration bit-strings during the evaluation of every $H_{ij}$ matrix element.  This saves memory but adds a substantial computational burden.  We took an alternative approach where all configuration bit-strings are computed one time and stored, and values are retrieved instead of being re-computed.  This requires an amount of memory equal to the total number of configurations times the size of each configuration bit-string, typically 8 to 32 bytes.  This optimization limits the total number of configurations to a value that depends on the available memory per device.  The GPUs on the Frontier supercomputer are AMD Instinct\texttrademark{} MI250X Accelerators with 64 GiB memory per graphics compute die (128 GiB per AMD Instinct MI250X Accelerator)~\cite{cdna2}. That limits the number of configurations to less than a few billion.  More recent GPUs have substantially larger memory capacities and could tackle somewhat larger problems.  The same optimization is also effective for CPUs, which may have up to a few TiB memory, and could handle even larger problems.  However, to reach or exceed ~$10^{12}$ configurations, one needs a different approach: either re-compute each configuration bit-string or devise a smaller cache that may require frequent updates.

In our measurements with the GPU-enabled code~\cite{amd_rocm_sys_profiler}, we found that performance was sensitive to the way the bit-strings are stored.  The CPU code~\cite{sqd-diag} provides a runtime option to set a ``bit\_length" parameter, which is given a default value of 20.  That means that at most 20 bits of information are set in each 64-bit \textit{size\_t} element.  Let's consider the case of the H$_2$O molecule with the cc-pvdz basis set.  There are 24 spatial orbitals, so 48 spin-orbitals.  If we use the default \textit{bit\_length = 20}, it would take three \textit{size\_t} elements for each configuration bit-string, 20 bits in each of the first two, and 8 bits in the third.  However, if we choose \textit{bit\_length = 48}, each configuration bit-string fits in a single \textit{size\_t} element.  This reduces the size of the configuration cache by a factor of three, and speeds up processing of the configuration bit-strings on the GPU.  In contrast, we found that the CPU code is not very sensitive to the choice of \textit{bit\_length}.

The cache uses OpenMP unstructured data management. Allocation occurs at initialization:

{\small
\begin{verbatim}
#pragma omp target enter data \
    map(alloc: cache[0:size])
\end{verbatim}
}
followed by parallel computation of all configurations:
{\small
\begin{verbatim}
#pragma omp target teams distribute \
    parallel for collapse(2)
for (ia = 0; ia < n_alpha; ia++)
  for (ib = 0; ib < n_beta; ib++)
    ComputeDet(&cache[offset], ...);
\end{verbatim}
}
The cache remains GPU-resident until program termination. This optimization saves computation at the expense of additional memory utilization.

\section{Modified data structures and GPU-specific routines}

\subsection{Flattening Nested Data Structures}

The CPU code's extensive use of nested \texttt{std::vector<std::vector<size\_t>>} structures posed the primary challenge for GPU porting. We systematically flattened all nested structures into three associated arrays. For excitation lists originally stored as \texttt{SinglesFromAlpha[ia][k]}, we transformed those to:

\begin{itemize}
    \item \texttt{SinglesFromAlpha\_flat}: concatenated excitation data
    \item \texttt{SinglesFromAlphaOffset[ia]}: starting index for configuration $ia$
    \item \texttt{SinglesFromAlphaLen[ia]}: number of excitations for configuration $ia$
\end{itemize}

The transformed memory access pattern:
\centerline{\texttt{SinglesFromAlpha\_flat[Offset[ia] + k]} }
provides single-indirection access with contiguous memory layout, enabling efficient coalesced GPU memory access. The flattening occurs once during helper construction and enables efficient data transfer using simple \texttt{map(to:)} clauses.
Similarly, two-electron integrals encapsulated in C++ classes with member functions were extracted into raw arrays. We replaced accessor functions like \texttt{I2.DirectValue(i,j)} with direct pointer arithmetic: \texttt{I2\_Direct[i + norbs $\times j$]}.

\subsection{Device-Side functions for evaluation of Hamiltonian matrix elements}

Evaluation of Hamiltonian matrix elements requires porting many quantum chemistry routines to GPU device code. Since OpenMP device code cannot use CPU intrinsics~\cite{openmp52} like \texttt{\_\_builtin\_ffsl()}, we implemented efficient, portable C++ equivalents.  The set of offloaded functions includes routines to get orbital occupancies, determine the parity or sign of each contribution, and specific functions to handle zero, one, or two excitations, according to Slater-Condon rules.  All functions are declared with \texttt{\#pragma omp declare target} for GPU compilation and placed in a single new source file, \texttt{hij\_omp\_offload.h}.

\subsection{GPU Memory Management Strategy}
A significant amount of data must reside on the GPU in order to evaluate matrix elements, including one and two electron integrals, lists of alpha and beta bit-strings, the configuration cache, and excitation lists that specify the sparse structure of the Hamiltonian matrix.  These data items are constant throughout the Davidson iterations, and could be copied to the GPU one time during program initialization.  In practice, the integral data and the configuration cache are transferred to the GPU one time using \texttt{\#pragma omp target enter data} directives.  In order to facilitate an incremental approach to offloading computational work, management of the excitation lists was moved into the matrix-vector multiply routine, mult().  The mult() routine also requires an input vector transferred to the GPU, and an output vector transferred to the GPU upon entry and returned to the CPU upon completion of each offload loop using OpenMP map(to:) and map(tofrom:) clauses respectively.  Recall that there are three distinct types of orbital differences that can contribute non-zero terms, and so there are three OpenMP work distribution loops, with \texttt{\#pragma omp target teams distribute parallel for} directives, and $collapse(2)$ clauses to ensure threading across both alpha and beta indices.  

Profiling measurements indicate that data transfer times are typically just a few percent of the total elapsed time or less, so no additional optimization of data movement is required (see Appendix B for details).

\section{Evaluation of the GPU enabled SQD algorithm}
Performance measurements were made using the Frontier supercomputer~\cite{frontier_supercomputer_ornl} at Oak Ridge National Laboratory.  Each node of Frontier has one CPU socket with a 64-core AMD EPYC\texttrademark{} 7A53 CPU, where 56 of the 64 cores are made available to the end-user (the other cores are used for system processes).  There are four MI250X GPU packages on each Frontier node, exposed as eight GPUs to the end user.  

We examined a large number of test cases (including a set of unit tests) with varying numbers of spin-orbitals and configurations, many different choices for the number of nodes or GPUs, and different MPI decomposition schemes.  The GPU code was verified to produce correct results (numerical correctness within $10^{-10}$ relative error) in all cases.  As a representative test case, we chose the H$_2$O molecule with the cc-pvdz basis and a set of bit-strings resulting in $6.28\times10^8$ configurations.  Timing measurements using 32 nodes of Frontier, 256 GPUs, are shown in Table~\ref{tab:32cpuvsgpu}.  

We note that the GPU code has one optimization that was not included in the original CPU code, namely the use of a persistent cache of configuration bit-strings.  Consequently we show performance relative to the original CPU code, and also relative to CPU code using the same configuration cache optimization.  All three measurements used the same number of MPI ranks and the same decomposition scheme, (8,8,4) with 8-way partitioning for each of the alpha and beta indices (64-way partitioning of the combined configuration space), and 4-way task parallelism. Each of the 256 MPI ranks has roughly ten million configurations to process.  The GPU code uses one MPI rank per GPU, and the CPU code uses OpenMP to engage all available CPU cores, including hyperthreads.

\begin{table}
    \caption{Measurements using 32 nodes of Frontier Supercomputer with different code versions using CPU and GPU hardware, and inputs for H$_2$O with $6.28\times10^8$ configurations.}
    \centering
    \begin{tabular}{|c|c|c|c|}\hline
         \textbf{Code Version} &  \textbf{Time(sec)}&  \textbf{Relative Perf }& \textbf{Relative Perf}\\
         & &\textbf{(CPU no cache)}& \textbf{(CPU with cache)}\\\hline
         CPU no cache&  5567.6&  1.00& 0.66x\\\hline
         CPU with cache&  3652.3&  1.52x& 1.00\\\hline
         GPU with cache&  58.3&  95x& 63x\\\hline
    \end{tabular}
    \label{tab:32cpuvsgpu}
\end{table}

The timing values listed in the table are the ``Davidson times" printed by the application.  This includes the time to compute and store diagonal terms, and the time spent in all iterations of the Davidson algorithm until the specified convergence criterion is met.  For our test case with $6.28\times10^8$ configurations, it takes a total of 11 iterations to converge.  There is one matrix-vector multiply per iteration, which consumes 96.5\% of the time in each iteration.  The remaining time is spent in the other stages of each Davidson iteration, including vector updates, dot-products, and Gram-Schmidt orthogonalization, where computation remains on CPUs.  There are also calls to MPI\_Allreduce to sum large arrays, and those take about 1\% of the elapsed time per iteration on Frontier for this test case.  If we take the original CPU code as our reference point, the GPU code is providing a ~95$\times$ speed-up on a per-node basis.  If we take CPU code with the configuration cache as our reference point (the same optimization used in the GPU code), the speed-up is ~63$\times$ on a per-node basis.  The configuration cache speeds up the original CPU code by a factor of ~1.52$\times$.

With the dramatic speed-up in computation, it is important to analyze the scaling behavior, with a focus on communication.  For the scaling test, we have chosen a slightly larger problem with $1.52\times10^9$ configurations, again using the H$_2$O molecule with the cc-pvdz basis set.  This is a quite large problem that would require many CPU hours.  Timing data for the GPU-accelerated code is shown in Table~\ref{tab:scaling}, from 1 node to 128 nodes, 8 to 1024 GPUs, along with the MPI decomposition schemes that were used.  The notation for MPI decomposition (a,b,t) indicates the number of ranks used to partition the alpha (a) and beta (b) bit-strings, and the number of ranks (t) used to concurrently process compute tasks.

\begin{table}
    \caption{Measurements for the GPU accelerated code on 1 node to 128 nodes, 8 to 1024 GPUs, using inputs for H$_2$O with $1.52\times10^9$ configurations on the Frontier Supercomputer.}
    \centering
    \begin{tabular}{|c|c|c|}\hline
         \textbf{Number of Nodes}&  \textbf{Time(sec)}& \textbf{Decomposition (a,b,t)}\\\hline
         1&  6384.7& (4,2,1)\\\hline
         2&  2606.2& (4,4,1)\\\hline
         4&  1057.2& (8,4,1)\\\hline
         8&  581.1& (8,8,1)\\\hline
         16&  318.0& (8,8,2)\\\hline
         32&  167.0& (8,8,4)\\\hline
         64&  99.9& (8,8,8)\\\hline
         128&  67.1& (8,8,16)\\\hline
    \end{tabular}
    \label{tab:scaling}
\end{table}

\begin{figure}[h!]
    \centering
    \includegraphics[width=0.9\linewidth, angle=90]{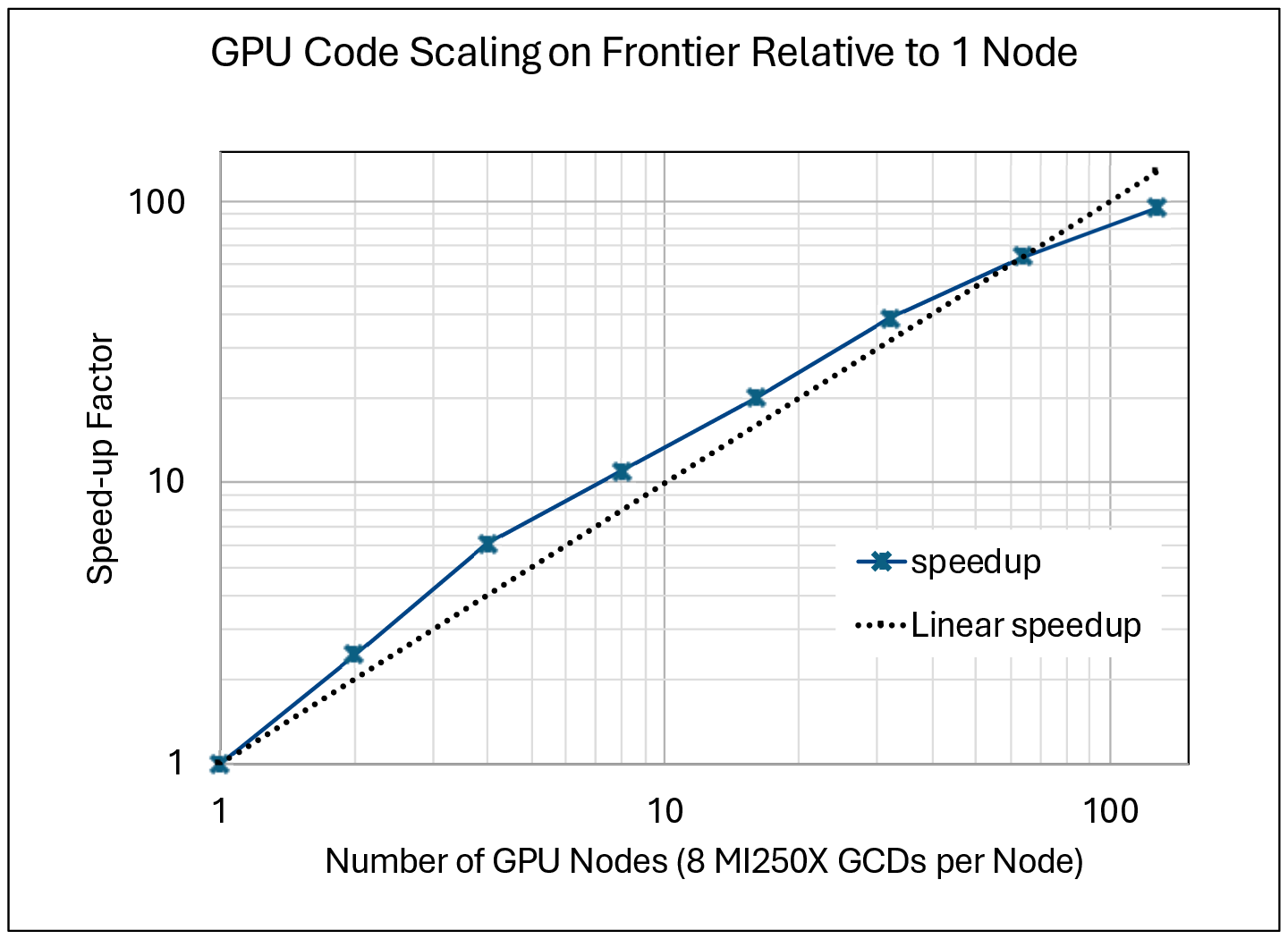}
    \caption{\bf{Strong scaling performance of GPU accelerated code on 1 to 128 nodes, 8 to 1024 GPUs on Frontier.}}
    \label{fig:scaling}
\end{figure}

For small node counts, we use only alpha and beta partitioning ($t=1$), and for larger node counts ($>=16)$ we turn on task-based parallelism ($t>1$).  These choices optimized the run time for a given node count.  As shown in Figure~\ref{fig:scaling}, for small node counts we observe super-linear speed-up, probably reflecting improved cache utilization as the number of configurations per GPU decreases.  For larger node counts the behavior is somewhat sub-linear, but parallel efficiency remains high ($\sim 75\%$) to at least 128 nodes, or 1024 GPUs.  

We used a MPI intercept library to collect detailed communication timing data, including time-line analysis.  Those measurements showed that parallel efficiency is mainly determined by load balance considerations, rather than bandwidth or latency in the network (see Appendix C for details).  When task-based parallelism is enabled, different task-groups take differing amounts of time to complete their work, resulting in reduced parallel efficiency.  For this test case at 128 nodes of Frontier, the network contribution remained less than 10\% of the elapsed time.

\section{Demonstrated portability}
\begin{figure}[h!]
    \centering
    \includegraphics[width=0.8\linewidth, angle=90]{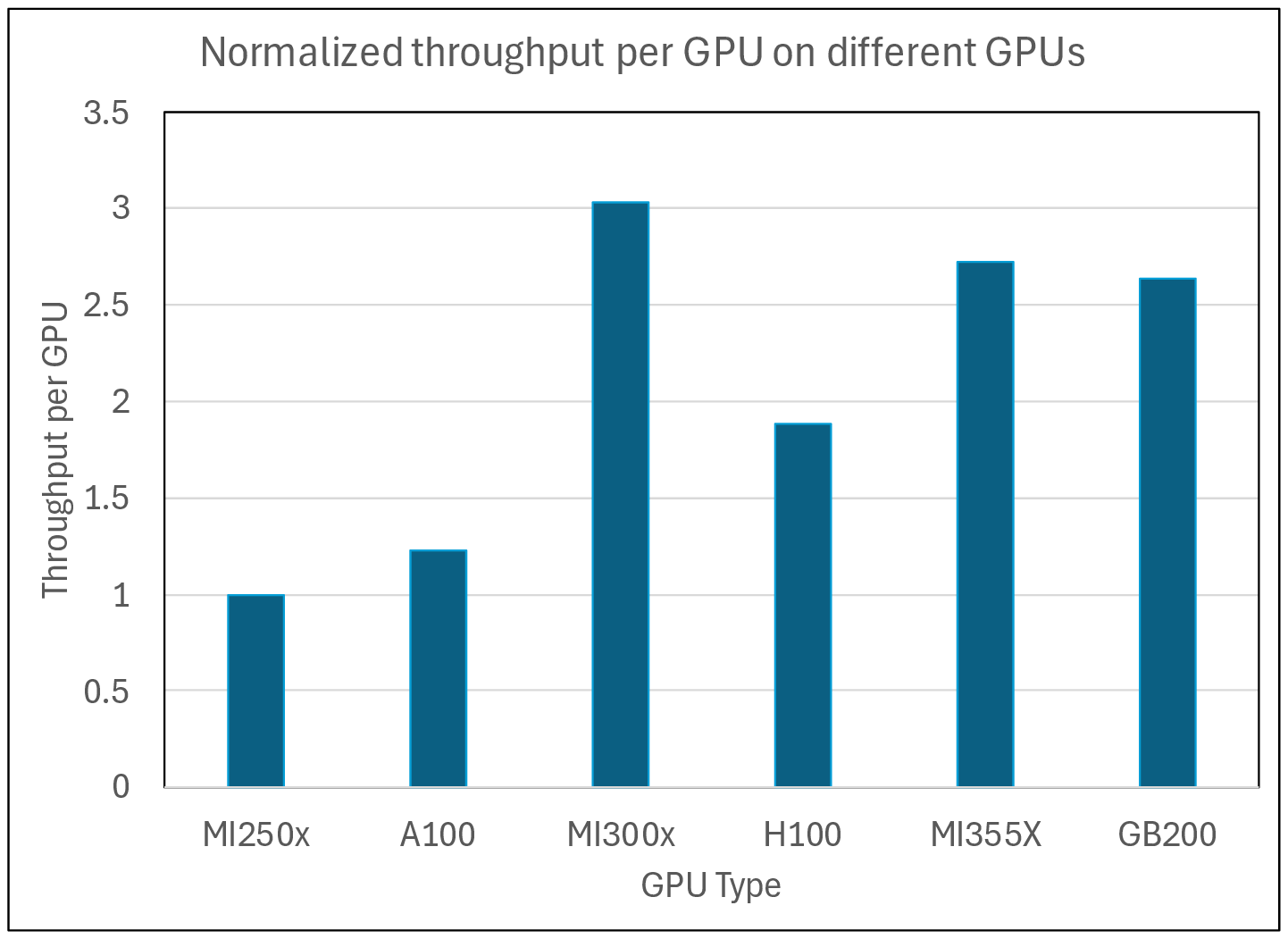}
    \caption{\bf{Comparison of performance per GPU across different GPU systems, using inputs for N$_2$ with $3.08\times10^8$ configurations problem. MI250X as the baseline. The figure shows three generations of GPUs: (MI250x, A100), (MI300X, H100), and (MI355X, GB200).}}
    \label{fig:thptgpu}
\end{figure}

We built and tested the OpenMP offload code on a variety of systems with x86 or Arm hosts, and AMD or NVIDIA GPUs.  No code changes were required; it was only necessary to specify correct architecture options for the compiler and re-compile.  Compilation was done using a version of the clang compiler: Cray clang version 20 on Frontier, and clang version 22 built from github source for the other systems.  These tests also provided an opportunity to see how the code performs on the latest GPU-accelerated systems.  Comparisons were made using just one or two nodes due to resource limitations.  Performance results using the $N_2$ molecule with $3.08\times10^8$ configurations are shown in Table~\ref{tab:gpus} and Figure~\ref{fig:thptgpu}, where the systems include Frontier with AMD MI250X~\cite{frontier_supercomputer_ornl}, a single-node system with eight AMD MI300X GPUs from IBM Cloud~\cite{ibm_gfpu_ai_accelerator_amd}, 
a single-node system with eight AMD MI355X GPUs, 
a single-node system with eight NVIDIA A100 GPUs, a system with eight NVIDIA H100 GPUs per node with Intel Sapphire Rapids CPUs~\cite{seelamvela} and Infiniband interconnect, and four Grace-Blackwell nodes with Arm CPUs and four NVIDIA GB200 GPUs per node.  Table entries show the times required for convergence and the performance per GPU, relative to an MI250X GPU (single GCD) on Frontier.  The newer GPUs are significantly faster, providing up to 3x more performance per GPU for AMD MI300X as shown in Figure~\ref{fig:thptgpu}.  For the family of AMD GPUs, performance approximately tracks the number of compute units (110 for MI250X, 304 for MI300X, and 256 for MI355X), with some additional dependence on frequency and caches.  This is consistent with our profiling analysis (see Appendix D for details), which indicated that performance depends mostly on scalar integer instruction throughput. MI300X is a balanced HPC and AI architecture. Continued HPC optimizations in GPUs are important for SQD and hybrid quantum-HPC workloads.

\begin{table}
    \centering
    \caption{Performance of diagonalization using inputs for N$_2$ with $3.08\times10^8$ configurations is shown for different GPU-accelerated systems. Performance per GPU uses MI250X as the baseline. The number of GPUs represents end-user visible devices.}
    \begin{tabular}{|c|c|c|c|}\hline
         \textbf{Accelerator}&  \textbf{Number of GPUs}&  \textbf{Time(sec)}& \textbf{Perf per GPU}\\\hline
         GB200&  16&  72.0& 2.64\\\hline
         MI355X&  8&  139.6& 2.72\\\hline        
         H100&  16&  100.7& 1.88\\\hline
         MI300X&  8&  125.1& 3.03\\\hline
         A100&   8&   309.0& 1.23\\\hline
         MI250X&  16&  189.8& 1.00\\\hline
    \end{tabular}
    \label{tab:gpus}
\end{table}

\section{Discussion}

The GPU-accelerated SQD implementation demonstrates that quantum chemistry codes with complex control flow and minimal floating-point intensity can nonetheless achieve substantial GPU acceleration. Despite the code exhibiting $\sim$55\% integer operations and only $\sim$1\% floating-point operations on CPU, the massive thread-level parallelism of GPUs effectively drives high instruction throughput.

Several factors contribute to the observed 95$\times$ speedup on a per-node basis. The persistent configuration cache eliminates redundant computation, accounting for approximately 1.5$\times$ improvement (as shown by CPU cache vs.\ no-cache measurements in Table~\ref{tab:32cpuvsgpu}). The remaining 60$\times$ speed-up stems from two factors: (1) there are eight GPUs per node compared to one CPU socket on Frontier, and (2) each MI250X graphics compute die provides 110 compute units (total of 220 compute units per MI250X accelerator) with 64 threads per unit (7,040 threads total), compared to 64 CPU cores with 2-way hyper-threading (128 threads).  This code places minimal demands on bandwidth to memory, and is characterized by an instruction mix dominated by scalar integer instructions.  The highly threaded architecture of GPUs enables a dramatic ~8x increase in instruction throughput per device compared to CPU sockets, provided there is sufficient parallelism to exploit.  In our case, this can be ensured by assigning at least ~$10^5$ configurations per GPU.

The OpenMP target offload approach proved effective for this porting effort, enabling portable code with minimal divergence between CPU and GPU paths. Conditional compilation via \texttt{\#ifdef USE\_GPU} maintains a single code base deployable on heterogeneous clusters. However, several limitations warrant discussion.  The GPU port needs a sufficient number of configurations per device in order to keep all compute units busy.  Problems with fewer than ~$10^5$ configurations can be solved efficiently on CPU systems.  In some cases it was convenient to set reasonable limits on the size of utility arrays, rather than relying on the more flexible methods supported by C++ \texttt{std::vector} objects.  Other considerations include:

\textbf{Memory capacity constraints:} The configuration cache requires $\mathcal{O}(n_\alpha \times n_\beta)$ GPU memory, limiting problems to $\sim$10$^9$ configurations on current MI250X hardware. Future work could explore hierarchical caching schemes with on-demand updates for memory-constrained regimes beyond $10^{10}$ configurations.

\textbf{Sensitivity to runtime parameters:} Sensitivity to the \textit{bit\_length} parameter on GPUs (but not CPUs) reveals architectural differences: GPUs benefit more from compact memory layouts due to limited cache capacity and higher memory latency. Setting \textit{bit\_length} to match actual spin-orbital count (e.g., 48 for H$_2$O/cc-pvdz) reduces cache size 3$\times$ and improves kernel performance, whereas CPU performance remains stable across \textit{bit\_length} choices (see Section~\ref{sec:cache}).  There is a shuffle option to randomize the sequence of bit-strings in the input.  This option is normally beneficial for the CPU code because it helps with load-balance, but the GPU version tends to be a little faster when the shuffle option is disabled. We made performance comparisons with the shuffle option enabled in all cases, which tends to favor the CPU code by a modest factor. The code provides four different ways to distribute work and memory.  This results in a large parameter space, making it challenging to find the best choices.  Currently exploring this space requires experimentation.  It would be useful to have more general guidance built into the program.

The phased optimization approach, with configuration caching, data flattening, careful data management, and incremental kernel porting, provides a reusable roadmap for similar GPU acceleration efforts in computational chemistry. Each phase delivers measurable performance improvements and can be independently validated, facilitating development and debugging.  By focusing on matrix-vector multiplication, our work can be immediately re-used in other iterative algorithms where matrix-vector multiplication plays a key role.

\section{Conclusions}

We have successfully ported the Sample-based Quantum Diagonalization method to GPU accelerators using OpenMP target offload directives, achieving 95$\times$ speedup on the Frontier supercomputer compared to CPU implementations. An additional 3$\times$ speedup was measured using the more recent MI300X GPU systems. The performance enhancement reduces diagonalization time from hours to minutes for representative molecular systems with up to $\sim$10$^9$ configurations, making SQD calculations tractable on moderately sized GPU resources.

The porting effort addressed fundamental challenges in GPU acceleration of quantum chemistry codes: transforming nested C++ data structures into flattened GPU-friendly arrays, implementing complete device-side evaluation of matrix elements according to Slater-Condon rules, and managing multi-tier GPU memory residency to minimize data transfers. The OpenMP target offload approach maintains a portable single-source code base functional on both CPU-only and GPU-accelerated systems, with conditional compilation enabling selective feature adoption.

Key technical contributions include: (1) a persistent GPU-resident configuration cache eliminating redundant computation, (2) systematic flattening of nested vectors into offset-indexed arrays for efficient memory access, and (3) complete GPU implementation of the quantum chemistry routines required to compute the Hamiltonian matrix, and (4) a portable solution that works across different GPU platforms.

Comprehensive validation across a wide range of test cases demonstrates numerical agreement with the CPU implementation to within a relative error of $10^{-10}$ for all excitation types. The code preserves all MPI decomposition strategies from the original CPU version, including partitioning of alpha and beta configurations, task-based parallelism, and row-communicator–based distribution.

\section*{Author Contributions}
The AMD team implemented the OpenMP offloading code. The IBM Research team conducted most experiments on AMD and NVIDIA systems reported in this paper. The ORNL team provided access to the Frontier system and all three teams contributed to writing and improving the manuscript.

\section*{Acknowledgments}
We gratefully acknowledge Shirakawa-san and our colleagues at RIKEN for their development of the highly scalable and performant CPU implementation of the subspace diagonalization subroutine for the SQD method, which provided an essential foundation for this work.

This research used resources of the Oak Ridge Leadership Computing Facility at the Oak Ridge National Laboratory, which is supported by the Office of Science of the U.S. Department of Energy under Contract No. DE-AC05-00OR22725.

AMD, the AMD Arrow logo, EPYC, Instinct, and combinations thereof are trademarks of Advanced Micro Devices, Inc.  Other product names used in this publication are for identification purposes only and may be trademarks of their respective companies.

\bibliographystyle{IEEEtran}
\bibliography{main}

@article{yu2025quantum,
  title={Quantum-Centric Algorithm for Sample-Based Krylov Diagonalization},
  author={Yu, Jeffery and Moreno, Javier Robledo and Iosue, Joseph T and Bertels, Luke and Claudino, Daniel and Fuller, Bryce and Groszkowski, Peter and Humble, Travis S and Jurcevic, Petar and Kirby, William and others},
  journal={arXiv preprint arXiv:2501.09702},
  year={2025}
}

@article{Huron_1973_CIPSI,
    author = {Huron, B. and Malrieu, J. P. and Rancurel, P.},
    title = {Iterative perturbation calculations of ground and excited state energies from multiconfigurational zeroth‐order wavefunctions},
    journal = {The Journal of Chemical Physics},
    volume = {58},
    number = {12},
    pages = {5745-5759},
    year = {1973},
    month = {06},
    issn = {0021-9606},
    doi = {10.1063/1.1679199},
    url = {https://doi.org/10.1063/1.1679199},
    eprint = {https://pubs.aip.org/aip/jcp/article-pdf/58/12/5745/18885418/5745_1_online.pdf},
}

@article{holmes2016efficient_SCI,
  title   = {Efficient heat-bath sampling in {F}ock space},
  author  = {Holmes, Adam A and Changlani, Hitesh J and Umrigar, CJ},
  journal = JCTC,
  volume  = {12},
  number  = {4},
  pages   = {1561--1571},
  year    = {2016},
  url     = {https://doi.org/10.1021/acs.jctc.5b01170}
}

@article{holmes2016heat_SCI,
  title   = {Heat-bath configuration interaction: An efficient selected configuration interaction algorithm inspired by heat-bath sampling},
  author  = {Holmes, Adam A and Tubman, Norm M and Umrigar, CJ},
  journal = JCTC,
  volume  = {12},
  number  = {8},
  pages   = {3674--3680},
  year    = {2016},
  url     = {https://pubs.acs.org/doi/10.1021/acs.jctc.6b00407}
}

@article{tubman2016deterministic_SCI,
  title   = {A deterministic alternative to the full configuration interaction quantum {Monte Carlo} method},
  author  = {Tubman, Norm M and Lee, Joonho and Takeshita, Tyler Y and Head-Gordon, Martin and Whaley, K Birgitta},
  journal = JCP,
  volume  = {145},
  number  = {4},
  year    = {2016},
  pages   = {044112},
  url     = {https://doi.org/10.1063/1.4955109}
}

@article{sharma2017semistochastic_SCI,
  title     = {Semistochastic heat-bath configuration interaction method: {S}elected configuration interaction with semistochastic perturbation theory},
  author    = {Sharma, Sandeep and Holmes, Adam A and Jeanmairet, Guillaume and Alavi, Ali and Umrigar, Cyrus J},
  journal   = JCTC,
  volume    = {13},
  number    = {4},
  pages     = {1595--1604},
  year      = {2017},
  publisher = {ACS Publications},
  url       = {https://pubs.acs.org/doi/10.1021/acs.jctc.6b01028}
}

@misc{zhang2025TrimCI_SCI,
      title={From Random Determinants to the Ground State}, 
      author={Hao Zhang and Matthew Otten},
      year={2025},
      eprint={2511.14734},
      archivePrefix={arXiv},
      primaryClass={quant-ph},
      url={https://arxiv.org/abs/2511.14734}, 
}

@misc{smith2025quantumcentricsimulationhydrogenabstraction_SQD,
      title={Quantum-centric simulation of hydrogen abstraction by sample-based quantum diagonalization and entanglement forging}, 
      author={Tyler Smith and Tanvi P. Gujarati and Mario Motta and Ben Link and Ieva Liepuoniute and Triet Friedhoff and Hiromichi Nishimura and Nam Nguyen and Kristen S. Williams and Javier Robledo Moreno and Caleb Johnson and Kevin J. Sung and Abdullah Ash Saki and Marna Kagele},
      year={2025},
      eprint={2508.08229},
      archivePrefix={arXiv},
      primaryClass={quant-ph},
      url={https://arxiv.org/abs/2508.08229}, 
}

@Article{Liepuoniute2025_SQD,
author={Liepuoniute, Ieva
and Doney, Kirstin D.
and Robledo Moreno, Javier
and Job, Joshua A.
and Friend, William S.
and Jones, Gavin O.},
title={Quantum-Centric Computational Study of Methylene Singlet and Triplet States},
journal={Journal of Chemical Theory and Computation},
year={2025},
month={May},
day={27},
publisher={American Chemical Society},
volume={21},
number={10},
pages={5062-5070},
issn={1549-9618},
doi={10.1021/acs.jctc.5c00075},
url={https://doi.org/10.1021/acs.jctc.5c00075}
}

@misc{sriluckshmy2025ghostgutzwiller_SQD,
      title={Quantum Assisted Ghost Gutzwiller Ansatz}, 
      author={P. V. Sriluckshmy and François Jamet and Fedor Šimkovic IV},
      year={2025},
      eprint={2506.21431},
      archivePrefix={arXiv},
      primaryClass={quant-ph},
      url={https://arxiv.org/abs/2506.21431}, 
}

@misc{shajan2026proteins_SQD,
      title={Molecular Quantum Computations on a Protein}, 
      author={Akhil Shajan and Danil Kaliakin and Fangchun Liang and Thaddeus Pellegrini and Hakan Doga and Subhamoy Bhowmik and Susanta Das and Antonio Mezzacapo and Mario Motta and Kenneth M. Merz Jr},
      year={2026},
      eprint={2512.17130},
      archivePrefix={arXiv},
      primaryClass={quant-ph},
      url={https://arxiv.org/abs/2512.17130}, 
}

@Article{Tubman2020,
author={Tubman, Norm M.
and Freeman, C. Daniel
and Levine, Daniel S.
and Hait, Diptarka
and Head-Gordon, Martin
and Whaley, K. Birgitta},
title={Modern Approaches to Exact Diagonalization and Selected Configuration Interaction with the Adaptive Sampling CI Method},
journal={Journal of Chemical Theory and Computation},
year={2020},
month={Apr},
day={14},
publisher={American Chemical Society},
volume={16},
number={4},
pages={2139-2159},
issn={1549-9618},
doi={10.1021/acs.jctc.8b00536},
url={https://doi.org/10.1021/acs.jctc.8b00536}
}

@article{Barison_2025_ext-SQD,
doi = {10.1088/2058-9565/adb781},
url = {https://doi.org/10.1088/2058-9565/adb781},
year = {2025},
month = {feb},
publisher = {IOP Publishing},
volume = {10},
number = {2},
pages = {025034},
author = {Barison, Stefano and Robledo Moreno, Javier and Motta, Mario},
title = {Quantum-centric computation of molecular excited states with extended sample-based quantum diagonalization},
journal = {Quantum Science and Technology}
}

@article{robledo2024chemistry,
author = {Javier Robledo-Moreno  and Mario Motta  and Holger Haas  and Ali Javadi-Abhari  and Petar Jurcevic  and William Kirby  and Simon Martiel  and Kunal Sharma  and Sandeep Sharma  and Tomonori Shirakawa  and Iskandar Sitdikov  and Rong-Yang Sun  and Kevin J. Sung  and Maika Takita  and Minh C. Tran  and Seiji Yunoki  and Antonio Mezzacapo },
title = {Chemistry beyond the scale of exact diagonalization on a quantum-centric supercomputer},
journal = {Science Advances},
volume = {11},
number = {25},
pages = {eadu9991},
year = {2025},
doi = {10.1126/sciadv.adu9991},
URL = {https://www.science.org/doi/abs/10.1126/sciadv.adu9991},
eprint = {https://www.science.org/doi/pdf/10.1126/sciadv.adu9991}}

@misc{sqd-diag,
  author = {Tomonori Shirakawa},
  title = {Libary for selected basis diagonalization},
  howpublished = {\url{https://github.com/r-ccs-cms/sbd}},
  year = {2025},
  note = {Accessed: 2025-10-06}
}

@article{kanno2023QSCI,
  title={Quantum-Selected Configuration Interaction: classical diagonalization of Hamiltonians in subspaces selected by quantum computers},
  author={Kanno, Keita and Kohda, Masaya and Imai, Ryosuke and Koh, Sho and Mitarai, Kosuke and Mizukami, Wataru and Nakagawa, Yuya O},
  journal={arXiv:2302.11320},
  year={2023},
  url={https://arxiv.org/abs/2302.11320}
}

@article{knowles1984new,
  title={A new determinant-based full configuration interaction method},
  author={Knowles, Peter J and Handy, Nicholas C},
  journal={Chemical physics letters},
  volume={111},
  number={4-5},
  pages={315--321},
  year={1984},
  publisher={Elsevier}
}

@article{davidson1975iterative,
title = {The iterative calculation of a few of the lowest eigenvalues and corresponding eigenvectors of large real-symmetric matrices},
journal = {Journal of Computational Physics},
volume = {17},
number = {1},
pages = {87-94},
year = {1975},
issn = {0021-9991},
doi = {https://doi.org/10.1016/0021-9991(75)90065-0},
url = {https://www.sciencedirect.com/science/article/pii/0021999175900650},
author = {Ernest R. Davidson}
}

@article{seelamvela,
  title={{The infrastructure powering IBM's Gen AI model development}},
  author={Gershon, Talia and Seelam, Seetharami and Belgodere, Brian and others},
  year={2025},
  eprint={2407.05467},
  archivePrefix={arXiv},
  primaryClass={cs.DC},
  url={https://arxiv.org/abs/2407.05467}, 
}

@article{Sun2020PySCF,
  title   = {Recent developments in the PySCF program package},
  author  = {Sun, Qiming and Berkelbach, Timothy C. and Blunt, Nick S. and Booth, George H.
             and Guo, Sheng and Li, Zhendong and Liu, Junzi and McClain, James D.
             and Sayfutyarova, Elvira R. and Sharma, Sandeep and Wouters, Sebastian
             and Chan, Garnet Kin-Lic},
  journal = {Journal of Chemical Physics},
  volume  = {153},
  number  = {2},
  pages   = {024109},
  year    = {2020},
  doi     = {10.1063/5.0006074}
}

@misc{shirakawa2025closedloopcalculationselectronicstructure,
      title={Closed-loop calculations of electronic structure on a quantum processor and a classical supercomputer at full scale}, 
      author={Tomonori Shirakawa and Javier Robledo-Moreno and Toshinari Itoko and Vinay Tripathi and Kento Ueda and Yukio Kawashima and Lukas Broers and William Kirby and Himadri Pathak and Hanhee Paik and Miwako Tsuji and Yuetsu Kodama and Mitsuhisa Sato and Constantinos Evangelinos and Seetharami Seelam and Robert Walkup and Seiji Yunoki and Mario Motta and Petar Jurcevic and Hiroshi Horii and Antonio Mezzacapo},
      year={2025},
      eprint={2511.00224},
      archivePrefix={arXiv},
      primaryClass={quant-ph},
      url={https://arxiv.org/abs/2511.00224}, 
}

@inproceedings{openmpoffloadoverview,
  title        = {OpenMP Target Offloading: Overview, Progress, and Future Directions},
  author       = {Valero-Lara, Pedro and Juckeland, Guido and Hernandez, Oscar},
  booktitle    = {IWOMP 2020: 16th International Workshop on OpenMP},
  year         = {2020},
  pages        = {1--15},
  doi          = {10.1007/978-3-030-58144-2_1}
}

@manual{nvidia_thrust_manual,
  title        = {Thrust: Parallel Algorithms Library},
  organization = {NVIDIA Corporation},
  year         = {2024},
  url          = {https://docs.nvidia.com/cuda/thrust},
}

@misc{hip,
  title        = {{HIP}: Heterogeneous-compute Interface for Portability},
  organization       = {{AMD}},
  year         = {2024},
  howpublished = {https://github.com/ROCm-Developer-Tools/HIP},
  note         = {Accessed: 2026-01-09}
}

@article{dagum1998openmp,
  title   = {{OpenMP}: An Industry-Standard API for Shared-Memory Programming},
  author  = {Dagum, Leonardo and Menon, Ramesh},
  journal = {IEEE Computational Science \& Engineering},
  volume  = {5},
  number  = {1},
  pages   = {46--55},
  year    = {1998},
  doi     = {10.1109/99.660313}
}

@misc{ibm_gfpu_ai_accelerator_amd,
  title        = {{AMD Instinct MI300X GPU on IBM Cloud}},
  organization       = {{IBM}},
  howpublished = {https://www.ibm.com/products/gpu-ai-accelerator/amd},
  note         = {Accessed: 2026-01-09},
  year         = {2026}
}

@misc{frontier_supercomputer_ornl,
  title        = {Frontier Supercomputer},
  organization       = {{Oak Ridge National Laboratory}},
  year         = {2022},
  url = {https://www.olcf.ornl.gov/frontier/},
  note         = {Accessed: 2026-01-09}
}

@misc{fugaku_supercomputer,
  title        = {Fugaku Supercomputer},
  organization       = {{Riken Center for Computational Science}},
  year         = {2021},
  url = {https://www.r-ccs.riken.jp/en/fugaku/},
  note         = {Accessed: 2026-01-09; Fugaku is a petascale supercomputer developed by RIKEN and Fujitsu, deployed at the RIKEN Center for Computational Science in Kobe, Japan. It was ranked the world’s fastest in 2020 and 2021 on the TOP500 list.} 
}

@misc{openmp52,
  title        = {{OpenMP} Application Programming Interface Version 5.2},
  organization = {OpenMP Architecture Review Board},
  year         = {2021},
  url          = {https://www.openmp.org/specifications/},
}

@manual{amd_rocm_sys_profiler,
  title        = {{ROCm} Systems Profiler Documentation},
  organization = {{AMD ROCm Documentation}},
  year         = {2025},
  note         = {Includes usage of tools such as \texttt{rocprof-sys-run}},
  url          = {https://rocm.docs.amd.com/projects/rocprofiler-systems/en/latest/},
}

@misc{perfetto_tracing_tool,
  title        = {Perfetto: System Profiling and Tracing Tool},
  organization = {Google},
  url = {https://perfetto.dev/},
  note         = {Open-source suite for system and application tracing and performance analysis},
  year         = {2025},
}

@misc{amd_uprof_performance_analysis,
  title        = {{AMD} $\mu$Prof Performance Analysis},
  organization       = {{Advanced Micro Devices, Inc.}},
  year         = {2025},
  url = {https://www.amd.com/en/developer/uprof/uprof-performance-analysis.html},
  note         = {Performance analysis tool for applications on AMD CPUs and GPUs},
}

@article{shajan2025molecular,
  title={Molecular Quantum Computations on a Protein},
  author={Shajan, Akhil and Kaliakin, Danil and Liang, Fangchun and Pellegrini, Thaddeus and Doga, Hakan and Bhowmik, Subhamoy and Das, Susanta and Mezzacapo, Antonio and Motta, Mario and Merz Jr, Kenneth M},
  journal={arXiv preprint arXiv:2512.17130},
  year={2025}
}

@article{piccinelli2025quantum,
  title={Quantum chemistry with provable convergence via randomized sample-based quantum diagonalization},
  author={Piccinelli, Samuele and Baiardi, Alberto and Rossmannek, Max and Vazquez, Almudena Carrera and Tacchino, Francesco and Mensa, Stefano and Altamura, Edoardo and Alavi, Ali and Motta, Mario and Robledo-Moreno, Javier and others},
  journal={arXiv preprint arXiv:2508.02578},
  year={2025}
}

@article{kaliakin2025accurate,
  title={Accurate quantum-centric simulations of supramolecular interactions},
  author={Kaliakin, Danil and Shajan, Akhil and Moreno, Javier Robledo and Li, Zhen and Mitra, Abhishek and Motta, Mario and Johnson, Caleb and Saki, Abdullah Ash and Das, Susanta and Sitdikov, Iskandar and others},
  journal={Research Square},
  pages={rs--3},
  year={2025}
}

@article{shajan2025toward,
  title={Toward quantum-centric simulations of extended molecules: Sample-based quantum diagonalization enhanced with density matrix embedding theory},
  author={Shajan, Akhil and Kaliakin, Danil and Mitra, Abhishek and Robledo Moreno, Javier and Li, Zhen and Motta, Mario and Johnson, Caleb and Saki, Abdullah Ash and Das, Susanta and Sitdikov, Iskandar and others},
  journal={Journal of Chemical Theory and Computation},
  volume={21},
  number={14},
  pages={6801--6810},
  year={2025},
  publisher={ACS Publications}
}

@manual{cdna2,
  title        = {Introducing {AMD CDNA}\texttrademark{} 2 Architecture},
  organization = {AMD},
  year         = {2021},
  url          = {https://www.amd.com/content/dam/amd/en/documents/instinct-business-docs/white-papers/amd-cdna2-white-paper.pdf},
}

@misc{horii,
      title={{GPU-Accelerated Selected Basis Diagonalization with Thrust for SQD-based Algorithms} }, 
      author={Doi, Jun and Shirakawa, Tomonori and Kawashima, Yukio and Horii, Hiroshi},
      year={2026},
      eprint={tbd},
      archivePrefix={arXiv},
      primaryClass={tbd},
      url={tbd}, 
}
\clearpage
\appendix
\section*{Appendix A: Hardware counter data for N$_2$ molecule}
\addcontentsline{toc}{section}{Appendix A: Hardware counter data for N2 molecule}
\label{hardwaredata}
\begin{table}[!h]
\centering
\caption{Performance Metrics Summary with CPU code using determinant cache.}
\label{tab:perf_metricscache}
\begin{tabular}{lll}
\hline
\textbf{Metric} & \textbf{Value} & \textbf{Notes} \\
\hline
Benchmark & N2 $6.67\times 10^6$ & CPU code with  \\
 &  &  determinant cache \\
Configuration & (4,2,1,1) & 12 threads \\
 &  & method = 0 \\
CPU & Intel Xeon &  \\
 & Platinum 8474C &  \\
\textbf{Elapsed Time} & \textbf{198 seconds} &  \\
\hline
\multicolumn{2}{c}{\textbf{Performance Metrics}} \\
\hline
perf::cycles & $6.09\times10^{11}$ & \textbf{IPC = 2.48} \\
perf::ref-cycles & $4.14\times10^{11}$ & FP = 1.2\% \\
perf::instructions & $1.51\times10^{12}$ & \textbf{LD\_ST = 26.2\%} \\
perf::branch-instructions & $2.68\times10^{11}$ & \textbf{Branch = 17.7\%} \\
MEM\_INST\_RETIRED:ALL\_LOADS & $2.76\times10^{11}$ & \textbf{Integer = 54.9\%} \\
MEM\_INST\_RETIRED:ALL\_STORES & $1.21\times10^{11}$ &  \\
FP\_ARITH\_INST\_RETIRED:SCALAR & $1.80\times10^{10}$ & max bw = 0.27 \\
FP\_ARITH\_INST\_RETIRED:VECTOR & $6.62\times10^{6}$ &  \\
FP\_ARITH\_INST\_RETIRED:4\_FLOPS & $6.62\times10^{6}$ &  \\
FP\_ARITH\_INST\_RETIRED:8\_FLOPS & $2.00$ &  \\
\hline
IMC Reads (per controller) & $\sim1.5\times10^{8}$ & imc0--imc7 \\
Read Bandwidth & 0.78 GB/s \\
IMC Writes (per controller) & $\sim5.6\times10^{7}$ & imc0--imc7 \\
Write Bandwidth & 0.29 GB/s \\
\hline
Measured DAXPY Bandwidth & 401 GB/s \\
\hline
\end{tabular}
\end{table}

\begin{table}[ht]
\centering
\caption{Performance Metrics Summary with CPU code no cache.}
\label{tab:perf_metricsnocache}
\begin{tabular}{lll}
\hline
\textbf{Metric} & \textbf{Value} & \textbf{Notes} \\
\hline
Benchmark & N2 $6.67\times 10^6$ & CPU code with  \\
 &  &  determinant cache \\
Configuration & (4,2,1,1) & 12 threads \\
 &  & method = 0 \\
CPU & Intel Xeon &  \\
 & Platinum 8474C &  \\
\textbf{Elapsed Time} & \textbf{314 seconds} \\
\hline
\multicolumn{2}{c}{\textbf{Performance Metrics}} \\
\hline
perf::cycles & \(9.66\times10^{11}\) & \textbf{IPC = 1.91} \\
perf::ref-cycles & \(6.55\times10^{11}\) & FP = 1.0\%\\
perf::instructions & \(1.84\times10^{12}\) & LD\_ST = 26.9\%\\
perf::branch-instructions & \(3.22\times10^{11}\) & \textbf{Branch = 17.5\%}\\
MEM\_INST\_RETIRED:ALL\_LOADS & \(3.55\times10^{11}\) & \textbf{Integer = 54.6\%}\\
MEM\_INST\_RETIRED:ALL\_STORES & \(1.41\times10^{11}\) \\
FP\_ARITH\_INST\_RETIRED:SCALAR & \(1.80\times10^{10}\) & max bw = 0.12\%\\
FP\_ARITH\_INST\_RETIRED:VECTOR & \(6.62\times10^{06}\) \\
FP\_ARITH\_INST\_RETIRED:4\_FLOPS & \(6.62\times10^{06}\) \\
FP\_ARITH\_INST\_RETIRED:8\_FLOPS & 2.00 \\
\hline
IMC Reads (per controller) & $\sim6.9\times10^{7}$ & imc0--imc7 \\
Read BW & 0.28 GB/s \\
IMC Writes (per controller) & $\sim4.8\times10^{8}$ & imc0--imc7 \\
Write BW & 0.20 GB/s \\
\hline
Measured DAXPY BW & 401 GB/s \\
\hline
\end{tabular}
\end{table}

Tables~\ref{tab:perf_metricscache} and ~\ref{tab:perf_metricsnocache} show the difference in IPC and memory access patterns with and without the determinant cache. With determinant cache, data is reused so there is an increase in read bandwidth, corresponding increase in IPC, and 50\% reduction in execution time from 314 seconds without cache to 198 seconds with cache.   

The overall summary for optimized CPU code was that each CPU core could complete $~2.5$ instructions per clock-cycle, where the instruction mix was $~55\%$ integer, $~26\%$ load-store, $~18\%$ branch, and only $~1\%$ floating-point, and where memory bandwidth utilization was negligible, at $~0.3\%$ of peak.

\clearpage
\section*{Appendix B: Sample profile of GPU kernels and memory operations}
\label{timeanalysis}
Table~\ref{tab:timeanalysis} shows aggregated profiling data over four GPUs for a representative test case. There are three OpenMP offload loops that take almost all of the elapsed time.  In this example, data transfer from host-to-device and device-to-host take a little over 1\% of the time.  Device-to-device memory transfers are associated with point-to-point communication and contribute very little to the overall elapsed time.  Similarly, there is one OpenMP offload evaluation of the determinant cache on each of the four GPUs, which takes a negligible fraction of the total elapsed time.

\begin{table}
    \centering
    \caption{The profiling measurements below report the percentage of time, number of calls, average time per call, and the corresponding operation. The operations memcpy H-to-D, D-to-H, and D-to-D represent the time associated with data movement from the host CPU to the GPU device, from the GPU device to the host CPU, and between GPU devices, respectively.}
    \begin{tabular}{|c|c|c|c|}\hline
         \textbf{Percent Time}&  \textbf{Call Count}&  \textbf{Avg. time(msec)}& \textbf{Operation}\\\hline
         69.9&  176&  119.7& omp loop 0\\\hline
         19.0&  88&  64.9& omp loop 1\\\hline
         9.9&  88&  34.0& omp loop 2\\\hline
         0.7&  4296&  0.049& memcpy H-to-D\\\hline
         0.5&  2464&  0.067& memcpy D-to-H\\\hline
         0&  1056&  0.004& memcpy D-to-D\\\hline
         0&  4&  0.443& omp det cache\\\hline
    \end{tabular}
    \label{tab:timeanalysis}
\end{table}

\section*{Appendix C: MPI time-lines from Frontier for a 32-node job}
\begin{figure}[h!]
    \centering
    \includegraphics[width=0.9\linewidth, angle=90]{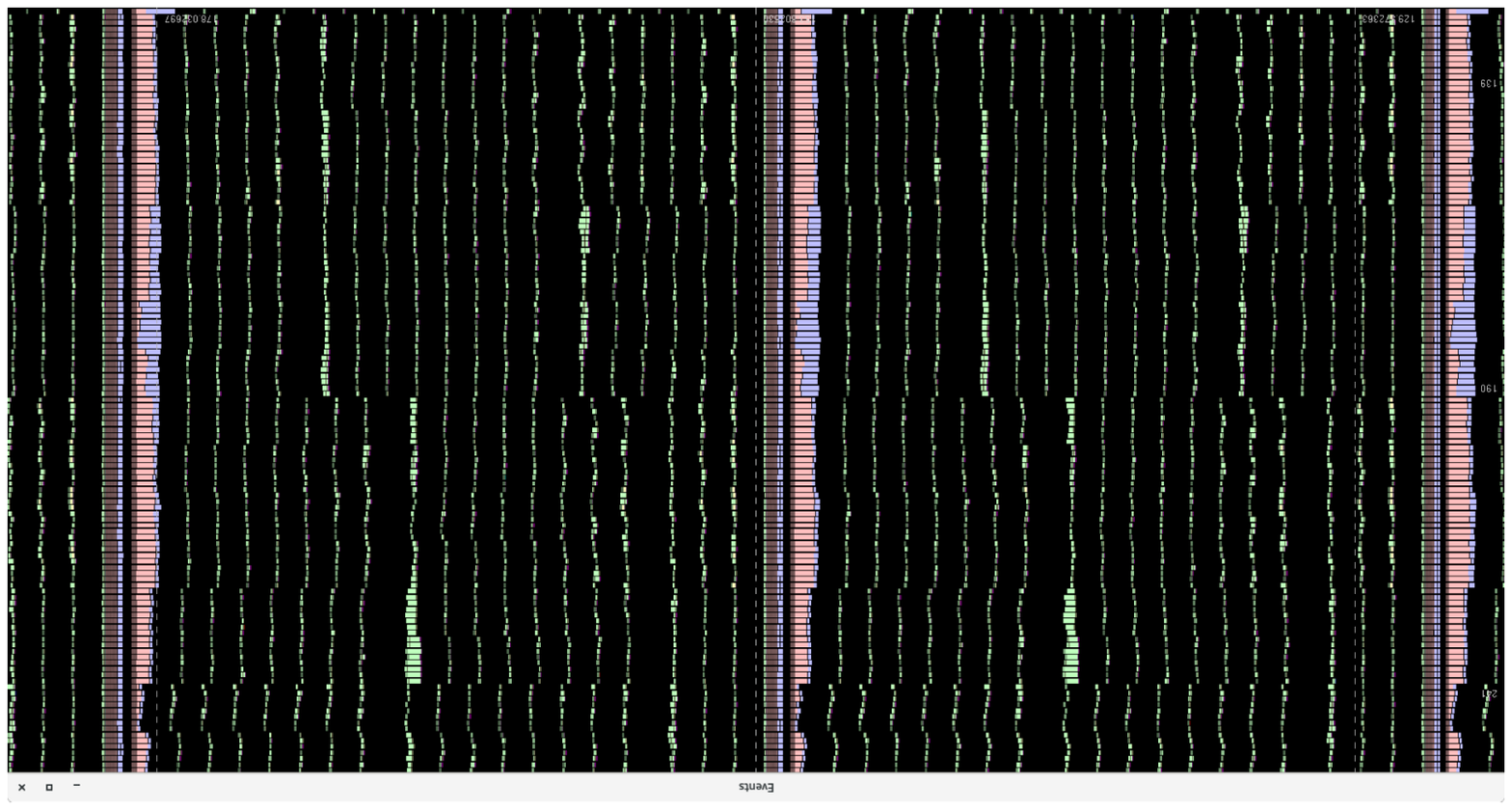}
    \caption{{\bf MPI time-lines for a representative job on 32 nodes of Frontier.}}
    \label{fig:mpi}
\end{figure}

\addcontentsline{toc}{section}{Appendix C: MPI time-lines from Frontier for a 32-node job}
\label{mpi}
Figure~\ref{fig:mpi} shows MPI time-lines from Frontier for a 32-node job with the H$_2$O inputs with $1.52*10^9$ determinants and (8,8,4,1) MPI decomposition. The y-axis is MPI rank (ranks 127-255 are shown) and the x-axis is time, showing two Davidson iterations. MPI events are colored boxes, black = computing, either on GPUs or CPUs. The light green boxes are from the point-to-point communication in Mpi2dSlide() during the mult() routine. The blue-ish boxes are calls to MPI\_Allreduce with large messages, and the orange-ish boxes are from MPI\_Allreduce with 8 byte sums. For this job a handful of MPI ranks near the top of the image take a little longer to do their computation, and other ranks have to wait in MPI for them. The small dark gap to the left of center-screen is due to vector updates on the CPU in the davidson() routine, then there is one call to MPI\_Allreduce with large messages followed by Gram-Schmidt orthogonalization, with a handful of calls to MPI\_Allreduce (8 byte sum) to complete the dot-products. This is using the original davidson() code from RIKEN~\cite{sqd-diag}. Overall, the performance remains compute-bound, and actual network traffic is not very important at this scale. Parallel efficiency is related more to load imbalance and/or differences in computation time, rather than network latency or bandwidth.

\clearpage
\section*{Appendix D: Hardware counter data for the NVIDIA GPUs using N$_2$ molecule using $1.92\times 10^6$ configurations}

The instruction mix shown in the tables below is very similar to what we measured with the CPU code, as shown in Appendix A. Instruction counts are basically the same for A100, H100, and GB200. There are three offload loops: type 2, 1, 0. We know from the structure of the code that loops of type 2 and type 1 should have the same instruction counts, and they do. But it takes nearly $2x$ longer to compute loop type 1. That is showing that data locality matters. Loop type 0 takes the most time. Frequency matters as well, GB200 is getting a good boost from frequency, and caches are helping on GB200, as evidenced by fewer bytes loaded from DRAM.

\begin{table}[htbp]
\centering
\caption{A100 Performance Metrics}
\label{tab:a100}
\begin{tabular}{llccccc}
\toprule
Metric & Unit & Type 2 & Type 1 & Type 0 & Sum & Percent \\
\midrule
Elapsed time & s & 0.048 & 0.093 & 0.292 & -- & -- \\
GPU cycles elapsed & cycles & 5.276E+07 & 1.032E+08 & 3.277E+08 & -- & -- \\
Effective frequency & cycles/s & 1.10E+09 & 1.11E+09 & 1.12E+09 & -- & -- \\
DRAM bytes read & MB & 927.92 & 49.62 & 171.98 & -- & -- \\
DRAM bytes written & MB & 57.53 & 9.94 & 36.30 & -- & -- \\
\midrule
\textbf{Thread instructions executed} & inst &
5.300E+11 & 5.289E+11 & 1.595E+12 & 2.65E+12 & \textbf{100.0} \\
\textbf{Integer instructions} & inst &
3.191E+11 & 3.184E+11 & 9.441E+11 & 1.58E+12 & \textbf{59.6} \\
\textbf{Control instructions} & inst &
1.437E+11 & 1.437E+11 & 4.633E+11 & 7.51E+11 & \textbf{28.3} \\
\textbf{FP64 instructions} & inst &
6.029E+09 & 6.029E+09 & 1.584E+10 & 2.79E+10 & \textbf{1.1} \\
\bottomrule
\end{tabular}
\end{table}

\begin{table}[htbp]
\centering
\caption{H100 Performance Metrics}
\label{tab:h100}
\begin{tabular}{llccccc}
\toprule
Metric & Unit & Type 2 & Type 1 & Type 0 & Sum & Percent \\
\midrule
Elapsed time & s & 0.0228 & 0.0444 & 0.1396 & -- & -- \\
GPU cycles elapsed & cycles & 4.539E+07 & 9.007E+07 & 2.868E+08 & -- & -- \\
Effective frequency & cycles/s & 1.99E+09 & 2.03E+09 & 2.05E+09 & -- & -- \\
DRAM bytes read & MB & 492.17 & 50.05 & 116.57 & -- & -- \\
DRAM bytes written & MB & 33.70 & 9.43 & 24.26 & -- & -- \\
\midrule
\textbf{Thread instructions executed} & inst &
5.496E+11 & 5.472E+11 & 1.656E+12 & 2.75E+12 & \textbf{100.0} \\
\textbf{Integer instructions} & inst &
3.160E+11 & 3.175E+11 & 9.523E+11 & 1.59E+12 & \textbf{57.6} \\
\textbf{Control instructions} & inst &
1.437E+11 & 1.437E+11 & 4.632E+11 & 7.51E+11 & \textbf{27.3} \\
\textbf{FP64 instructions} & inst &
6.029E+09 & 6.029E+09 & 1.584E+10 & 2.79E+10 & \textbf{1.0} \\
\bottomrule
\end{tabular}
\end{table}

\begin{table}[htbp]
\centering
\caption{GB200 Performance Metrics}
\label{tab:gb200}
\begin{tabular}{llccccc}
\toprule
Metric & Unit & Type 2 & Type 1 & Type 0 & Sum & Percent \\
\midrule
Elapsed time & s & 0.0192 & 0.0376 & 0.1205 & -- & -- \\
GPU cycles elapsed & cycles & 5.727E+07 & 1.136E+08 & 3.668E+08 & -- & -- \\
Effective frequency & cycles/s & 2.98E+09 & 3.02E+09 & 3.04E+09 & -- & -- \\
DRAM bytes read & MB & 49.87 & 49.51 & 47.86 & -- & -- \\
DRAM bytes written & MB & 0.256 & 0.799 & 5.38 & -- & -- \\
\midrule
\textbf{Thread instructions executed} & inst &
5.475E+11 & 5.450E+11 & 1.651E+12 & 2.74E+12 & \textbf{100.0} \\
\textbf{Integer instructions} & inst &
3.154E+11 & 3.174E+11 & 9.523E+11 & 1.59E+12 & \textbf{57.8} \\
\textbf{Control instructions} & inst &
1.448E+11 & 1.448E+11 & 4.667E+11 & 7.56E+11 & \textbf{27.6} \\
\textbf{FP64 instructions} & inst &
6.029E+09 & 6.029E+09 & 1.584E+10 & 2.79E+10 & \textbf{1.0} \\
\bottomrule
\end{tabular}
\end{table}

\end{document}